\documentclass[10pt]{article} 
\usepackage{cite}
\usepackage{amsmath} 
\usepackage{amssymb}
\usepackage{epsfig}
\usepackage{cite}
\usepackage{graphics}
\begin{document}

\begin{center} 
{\Large \bf Homogeneous nucleation near a second phase transition
and Ostwald's step rule} \vskip 0.3cm

{Z.~Tavassoli \footnote{Electronic mail address:
z.tavassoli@surrey.ac.uk} and R.~P.~Sear  \footnote{Electronic mail address:
r.sear@surrey.ac.uk}}

\vskip 0.1cm {\it Department of Physics, University of Surrey, Guildford,
Surrey, GU2 7XH, U.K.}

\end{center}

{\noindent \large \bf Abstract}

\noindent
Homogeneous nucleation of the new phase of one transition near a
second phase transition is considered. The system has two phase
transitions, we study the nucleation of the new phase of one of these
transitions under conditions such that we are near or at the second
phase transition. The second transition is an Ising-like transition
and lies within the coexistence region of the first transition.
It effects the formation of the new phase in two
ways. The first is by reducing the nucleation barrier to direct
nucleation. The second is by the system undergoing the second
transition and transforming to a state in which the barrier to
nucleation is greatly reduced. The second way occurs when the barrier
to undergoing the second phase transition is less than that of the
first phase transition, and is in accordance with Ostwald's rule.

%\vskip 0.15cm \noindent%

%\vskip 0.2cm \noindent Keywords:

%\noindent PACS numbers:

%\newpage

\section{Introduction} \label{introduction}

The formation of a new phase at a first-order phase transition is an
activated process. A nucleus of the new phase must overcome a
free energy barrier in order to form and then grow into the
new phase. The rate at which such microscopic nuclei form is
proportional to $\exp(-\Delta F^*/kT)$, where $\Delta F^*$ is the
height of the free energy barrier which the nuclei must overcome
\cite{Deb96}.
Here we consider nucleation of the new phase at a first-order
phase transition and calculate $\Delta F^*$ when the system is near a
second phase transition.
One of the examples of nucleation near a second transition is
nucleation of a crystalline phase at a first-order fluid-crystal
transition near a fluid-fluid phase transition.
Such a process has been observed in globular proteins whose phase
diagrams show a metastable fluid-fluid transition within a strongly
first-order fluid-crystal transition \cite{BBPOB91,MR97}.
Crystallisation of globular proteins is the subject of interest as
protein crystals are needed in order to study their structure by
X-ray diffraction \cite{RVMT96,CH98,Piaz00}.
It has been seen that globular proteins crystallize at temperatures
near where we expect a metastable fluid-fluid critical point
\cite{GW94,RZZ96}.
Numerical work on nucleation near a critical point has been done by
Talanquer and Oxtoby \cite{TO98}, where they obtained nuclei with
very large numbers of molecules near the critical point.
This followed pioneering computer simulations of ten Wolde and
Frenkel \cite{tWF97}, who found an anomalously low $\Delta F^*$ for
the nucleation of a crystalline phase near the critical point of a
metastable fluid-fluid transition.
Other recent theoretical work on nucleation near a metastable
transition may be found in
\cite{HD00,DZ00,PJPR00,Sear01a,Sear01b,Sear01c} and references
therein. Earlier theoretical work by one of us
\cite{Sear01a,Sear01b,Sear01c} showed that as the critical point is
approached, the derivatives of the
free energy barrier to nucleation with respect to the
chemical potential and temperature diverge.
Therefore the presence of the critical point causes a rapid drop in
the free energy barrier to nucleation and so facilitates nucleation.

The metastable transition we consider here is an Ising-like or a
vapour-liquid-like transition.
An Ising-like transition is a transition from a phase with a negative
magnetisation to a phase with a positive magnetisation.
A vapour-liquid-like transition is a transition
between two fluid phases differing in density.
The metastable transition lies within the coexistence region of the
equilibrium transition. Thus when the system is in the coexistence
region of the metastable phase transition, it also lies within the
coexistence region of the equilibrium transition.
Then one of two new phases can nucleate: the stable phase can
nucleate, as can the liquid phase or
the analogue of the liquid phase, from the vapour phase or the
analogue of the vapour phase.
Which of these occurs first depends on which free energy barrier is
lower.
Thus the stable phase can be formed via two
processes. In one process the free energy barrier to nucleation of
the stable phase becomes small enough to allow nucleation, but the
barrier to formation of the liquid phase
is still large, then the stable phase nucleates.
When the sizes of the barriers are reversed, then the formation of
the stable phase occurs via two steps: first the liquid phase
nucleates, and second the stable phase nucleates in this liquid
phase.

Our finding that nucleation of a metastable transition can occur
before nucleation of the stable phase, is just that observed by
Ostwald more than $100$ years ago \cite{Ost97,tWF99}.
Ostwald's rule is that the phase that nucleates first is not
necessarily the most thermodynamically stable phase, but is the one
with closest free energy to the fluid phase. Later Stranski and
Totomanow \cite{ST33} improved this rule and suggested that the
nucleated phase is the one with the lowest free energy barrier to
nucleation, and not the phase that is globally stable.

Nucleation can be heterogeneous or homogeneous.
In most circumstances suspended and dissolved impurities, as well as
the solid boundaries provide sites for the formation of the new
phase. This process is known as heterogeneous nucleation.
However, in the absence of impurities or solid surfaces, small nuclei
of the new phase are formed within the bulk of the system.
This is a homogeneous nucleation and
this is what we are studying here.
A free energy barrier must be overcome in order to form nuclei of a
critical size, beyond which the new phase grows spontaneously
\cite{Deb96,Abr74}. The rate at which critical nuclei of the new
phase are formed is very sensitive to the height of the free energy
barrier. The so-called classical nucleation theory was originated
with the work of Volmer and Weber \cite{VW26,Abr74} and
originally developed for droplet condensation from supercooled
vapours. The theory calculates the rate at which nuclei grow to a
critical size, proportional to $\exp(-\Delta F^*_I/kT)$,
where $\Delta F^*_I$ is the minimum work needed to form the critical
nucleus.

In this paper we carry out numerical calculations to study the
homogeneous nucleation of the new phase at a first-order
transition near or at the second phase transition.
The second transition is an Ising-like transition from the phase
with a negative magnetisation to the phase with a positive
magnetisation. In this work we use magnetic language, in
fluid language the transition is from a low density phase to a high
density phase.
We present the results of numerical calculations for the
temperatures below the critical point.
The calculations are for the excess order parameter
$\Delta m^*$ and the free energy barrier to nucleation of the new
stable phase, $\Delta F^*$, in the critical nucleus.
In the next section we study the theory of two types of homogeneous
nucleation: nucleation of the new stable phase, and nucleation
of the positive magnetisation phase from the negative magnetisation
phase.
We then discuss some of the numerical techniques we used in
the calculations in section \ref{numerical}. In section
\ref{results} the results are presented and the last section is a
conclusion.

%\newpage

\section{Theory}    \label{theory}

\subsection{Nucleation of a stable phase}

We study nucleation of a new phase at a first-order transition
near a second transition. The second transition is an Ising-like
transition with an order parameter $m$. We work below the critical or
Curie temperature and so we have a transition from negative $m$ to
positive $m$ on increasing the field $h$. $h$ is as usual the field
which couples to $m$. Following
earlier work by one of us \cite{Sear01a} we split the nucleus of the
new phase into two parts: the core and the fringe. That part of
nucleus less than $r_c$ from the origin is the core and the part
farther than $r_c$ is the fringe of the nucleus. We assume that the
fringe of the nucleus is spherically symmetric. Therefore the order
parameter $m(r)$ is a function only of $r$, the distance from the
centre of the nucleus. As the universal behaviour of the nucleus is
derived from the fringe, we therefore substitute the core by a
boundary condition on $m(r)$ of the fringe. Thus for simplicity we
fix $m(r \le r_c) = m_c$, where the subscript $c$ stands for
the core of the nucleus. $m_c$ is a positive constant independent of
temperature and of $h$, and is essentially the value of the order
parameter $m$ in the nucleating phase. We require the excess free
energy of the nucleus, $\Delta F$, which is the
free energy with a nucleus present minus that without a nucleus. Here
we use a standard square-gradient functional for the fringe of the
nucleus as a functional of the order parameter profile of the
nucleus, $m(r)$, \cite{Deb96,CH59,CH58,Wid85}

\begin{equation} \label{paper1-1} 
\Delta F =  \Delta F_c +
\int_{r>r_c} \left[ \Delta\omega(m) + \kappa \left(\nabla m\right)^2 \right] d{\bf r},
\end{equation}
%%%%%%%%%%%%%%%%%%%%%%%
where the first term is
the contribution from the core of the nucleus, $\Delta F_c =
\frac{4}{3} \pi r_c^3 \Delta \omega_{core}$, and the integral term
is the contribution from the fringe of the nucleus.
In Eq.~(\ref{paper1-1}) and all other equations, our free energies
are in units of $kT$; $k$ and $T$, Boltzmann's constant and the
temperature, respectively. $\Delta \omega_{core}$ is a constant and
is the excess free energy per unit volume of the core. As the core is
close to the bulk equilibrium phase, $\Delta \omega_{core} < 0$
because the free energy is lower in the equilibrium phase.
The gradient term in the integrand of Eq.~(\ref{paper1-1}) is due to
the variations in space of $m(r)$. This is the lowest order term in a
gradient expansion and is adequate when $m$ is slowly varying. The
coefficient $\kappa$ is assumed to be a constant. In
Eq.~(\ref{paper1-1}) $\Delta \omega$ in the integrand is given by

\begin{equation}  \label{paper1-2}
\Delta \omega = f(m)-f(m_b)-h(m-m_b), 
\end{equation}
where $m_b$ is the order parameter in the bulk.
$\Delta \omega$ is the work required per unit volume to change the
order parameter from its bulk value
$m_b$ to $m$ in the presence of an external field $h$. $f(m)$ is the
bulk Helmholtz free energy per unit volume as a function of the order
parameter $m(r)$. The critical nucleus is at the top of the
free energy barrier and therefore at the maximum of
$\Delta F$. So we set the functional derivative of $\Delta F$ with respect to the order parameter profile $m(r)$, to zero,

\begin{equation}  \label{paper1-8}
\frac{\partial \Delta \omega(m)}{\partial m(r)} - 2\kappa \nabla^2 m(r) =
 0,    \quad r \geq r_c,
\end{equation}
%%%%%%%%%%%%%%%%%%%%%%
The total excess of the order parameter due to the presence of the
nucleus, which is the total order parameter with the nucleus
present minus that without the nucleus, is obtained as

\begin{equation}  \label{paper1-8.5}
\Delta m = \int (m(r)-m_b) d{\bf r}.
\end{equation}
%%%%%%%%%%%%%%%%%

We employ Landau theory \cite{Kad00,CL95} for
the bulk free energy,

\begin{equation}  \label{paper1-4}
f(m,T) = \frac{1}{2}atm^2 + b m^4,
\end{equation}
where $a$ and $b$ are constants, and $t=T-T_c$ with $T_c$ the
critical temperature \cite{CL95}. The bulk order parameter $m_b(h,T)$
is obtained by solving

\begin{equation} \label{paper1-5}
\frac{\partial \omega}{\partial m} \bigg|_{m=m_b} = 0,
\end{equation}
%%%%%%%%%%%%%%%%%%%%%%%%
where $\omega(m)=f(m)-hm$. Then using
Eqs.~(\ref{paper1-4},\ref{paper1-5}) gives

\begin{equation}  \label{paper1-7}
atm + 4bm^3 - h = 0.
\end{equation}
%%%%%%%%%%%%%%%%%%
When $h=0$ the bulk order parameter is that at coexistence and we
call it $m_{co}$. For $h=0$, Eq.~(\ref{paper1-7})
has just one real solution $m_{co}=0$ for $T>T_c$.
For $h=0$ and $T<T_c$, Eq.~(\ref{paper1-7}) has three
solutions, and consequently the free energy has two minima with
the same free energy at
$m_{co}=\pm \frac{1}{2}\left(-at/b\right)^{1/2}$, and a maximum at
$m=0$.
In the presence of an external field $h$, Eq.~(\ref{paper1-7}) has
one real solution for $T>T_c$, which corresponds to a single minimum
for $\omega(m)$. When $T<T_c$ and $h \ne 0$ with $|h|$ not too large,
there are two inequivalent minima in $\omega (m)$.
For a fixed $h$, the absolute minimum corresponds to
the equilibrium state and the higher minimum is a metastable state.

To get the order parameter profile we substitute Eq.~(\ref{paper1-4})
in Eq.~(\ref{paper1-8}), which gives

\begin{equation}  \label{paper1-9}
\frac{d^2m}{dr^2} + \frac{2}{r} \frac{dm}{dr} - \frac{1}{2\kappa} \left( 4bm^3 + atm -h 
\right) = 
0,
\end{equation}
for $r>r_c$, subject to boundary conditions

\begin{equation}  \label{paper1-10}
m(r=r_c)=m_c
\end{equation} 
and

\begin{equation}  \label{paper1-11}
m(r\to\infty)=m_b.
\end{equation}
The former is the order parameter at the boundary between the fringe
and the core, and the latter is the obvious condition that $m$ tends
towards its bulk value far from the nucleus. Eq.~(\ref{paper1-9}) is
a nonlinear second-order differential equation which can not be
solved analytically.
We therefore use numerical methods to solve Eq.~(\ref{paper1-9}) for
the order parameter profile, and then substitute it in
Eq.~(\ref{paper1-1}) to obtain the excess free energy of the
nucleus. The barrier to nucleation is at the maximum of the excess
free energy and so will occur at a radius $r^*_c$ given by $\partial
\Delta F/\partial {r_c} = 0$. Thus we need to use a maximization
method numerically to find $r^*_c$ and hence $\Delta F^*$, which is
the free energy barrier to nucleation for a critical nucleus.
This also gives $m^*(r)$ which is the order parameter profile of the
critical nucleus.
Substituting this profile in
Eq.~(\ref{paper1-8.5}) we then find the excess order parameter of
the critical nucleus $\Delta m^*$.

\subsection{Nucleation of $+m$ phase from $-m$ phase, Classical
 Nucleation Theory}

Starting in the $-m$ phase and increasing $m$ (or equivalently $h$),
we cross the phase transition from the $-m$ to the $+m$ phase. Then
our system can not only nucleate the equilibrium phase but also the
$+m$ phase. Then we are within both the $-m$ to $+m$ coexistence
region and the coexistence region between the equilibrium phases.
The nucleus of the $+m$ phase must overcome a free energy barrier in
order to form, and then it grows to the new phase.
The free energy barrier to form such critical nucleus is given by

\begin{equation}  \label{paper1-n14}
\Delta F^*_I = \frac{16 \pi \sigma^3}{3h^2},
\end{equation}
%%%%%%%%%%%%%%%%%%%%%%
which is the standard classical nucleation theory expression
\cite{Deb96,CL95,LL80}.
This expression is valid for nucleation not too
close to the spinodal \cite{Bin84}.
The subscript $I$ indicates that the transition is an Ising-like
phase transition.
In Eq.~(\ref{paper1-n14}) $\sigma$ is the
surface free energy of the equilibrium interface between $+m$ and
$-m$ phases, and is given by

\begin{equation} \label{paper1-n15}
\sigma = 2 \int_{-m_{co}}^{m_{co}} \sqrt{\kappa \Delta f(m)} dm,
\end{equation}
%%%%%%%%%%%%%%%%%%%
where $\Delta f(m) = f(m) - f(m_b)$ is the excess bulk free energy,
Eq.~(\ref{paper1-4}).

It is worth mentioning that in this paper we have used a magnetic
rather than a fluid language to describe the phase transitions.
To switch from the magnetic language with a transition from $-m$ to
$+m$ phase, to the fluid language with a vapour-liquid-like
transition, the external field $h$ becomes the chemical potential
$\mu$ minus that at the coexistence $\mu_{co}$, the order parameter
$m$ becomes density $\rho$, the bulk order parameters $-m_b$ and
$+m_b$ become the vapour and liquid densities, $\rho_v$ and $\rho_l$,
respectively, and the axis $m=0$ becomes the critical density
$\rho_{cp} = (\rho_v+\rho_l)/2$ in the coexistence curve. Also in the
fluid language the total order parameter $\Delta m^*$ is equivalent
to the total number of molecules in the critical nucleus.

\section{Numerical Techniques}   \label{numerical}

In the fringe far from the core, $m(r)$ is near the bulk order
parameter $m_b$, we therefore use a Taylor expansion of $\Delta
\omega(m)$ about $m=m_b$,

\begin{equation}  \label{paper1-n1}
\Delta \omega(m) = \frac{1}{2} \chi^{-1}(m-m_b)^2+ \cdots
\end{equation}

\begin{equation} \label{paper1-n2}
\frac{\partial \Delta \omega(m)}{\partial m} = \chi^{-1}(m-m_b)+ \cdots,
\end{equation}
where $\Delta \omega$ and its first derivative are zero at $m=m_b$.
Here $\chi$ is the response function of $m$, defined by

\begin{equation}  \label{paper1-n3}
\chi^{-1} = \frac{\partial^2 f}{\partial m^2} \bigg|_{m=m_b},
\end{equation}
where $f(m)$ is the bulk free energy Eq.~(\ref{paper1-4}).
Substituting Eq.~(\ref{paper1-n2}) in Eq.~(\ref{paper1-8}) gives

\begin{equation} \label{paper1-n4}
\chi^{-1}(m-m_b) - 2\kappa \nabla^2 (m-m_b) = 0,
\end{equation}
which has a solution of Yukawa form

\begin{equation} \label{paper1-n5}
m(r)=m_b + B\,\frac{\exp(-r/\xi)}{r},
\end{equation}
for large values of $r$.
To obtain Eq.~(\ref{paper1-n5}) we used the boundary condition
Eq.~(\ref{paper1-11}). In Eq.~(\ref{paper1-n5}) $B$ is a constant
value which we will fix later, and $\xi$ is the correlation length
for $m(r)$ which is given by

\begin{equation}  \label{paper1-n6}
\xi = \left(2\kappa \chi \right)^{1/2}.
\end{equation}
%%%%%%%%%%%%%%%%%%%%%%%%
The solution of $m(r)$ for large values of $r$ has the form of
Eq.~(\ref{paper1-n5}). In numerical calculations we therefore use a
large distance from the center of the core, $R$,
to be the outer boundary of our numerical integration of
Eq.~(\ref{paper1-9}), and use Eq.~(\ref{paper1-n5}) beyond $R$.

For our numerical calculations we employ the Runge-Kutta method to
solve the second-order differential Eq.~(\ref{paper1-9}) for
the order parameter profile, subject to boundary conditions
at $r_c$ and $R$, Eq.~(\ref{paper1-10}) and Eq.~(\ref{paper1-n5}),
respectively. The constant value $B$ in Eq.~(\ref{paper1-n5}) is then
obtained by finding the root of the nonlinear equation
$m(r_c)-m_c=0$, for which we used Ridders' method \cite{PTVF92}.
Substituting the order parameter profile $m(r)$ in
Eq.~(\ref{paper1-1}) gives the excess free energy of the nucleus.
The barrier to nucleation is at a
maximum of $\Delta F$ and so will occur at a radius $r_c^*$ given
by $\partial \Delta F/\partial r_c = 0$.
We then use a maximization method to obtain $r_c^*$ and consequently
$m^*(r)$ and the barrier to nucleation $\Delta F^*$.

In the numerical calculations we used a large radius $R$ for the
boundary condition of the fringe of the nucleus. We need such a value
to enable us to carry out the numerical calculations. We calculate
the contribution of distances larger than $R$ analytically and add it
to the corresponding numerical quantity for the range $r_c$ to $R$,
as a correction term. The correction term for the excess order
parameter of the critical nucleus, $\Delta m^*$, can be calculated
using Eq.~(\ref{paper1-8.5}) with the limits for the integral from
$R$ to $\infty$. Then by substituting Eq.~(\ref{paper1-n5}) for
$m(r)$ for all values of $r \geq R$, the correction term is

\begin{equation}  \label{paper1-n9}
\Delta m^*_{ct} = 4\pi \kappa B \xi (R+\xi) e^{-R/\xi},
\end{equation}
where $ct$ stands for the correction term.
The correction term for the free energy barrier to nucleation
$\Delta F^*$, is calculated using the integral term in
Eq.~(\ref{paper1-1}) for all values of $r$ from $R$ to $\infty$.
Then by substituting Eq.~(\ref{paper1-n1}) in $\Delta \omega(m)$
and using the normal gradient of Eq.~(\ref{paper1-n5}) for $\nabla m(r)$, the correction term is given by

\begin{equation}  \label{paper1-n10}
\Delta F^*_{ct} = 4\pi \kappa B^2 \left(\frac{1}{R} + \frac{1}{\xi} \right) e^{-2R/\xi},
\end{equation}
which we have used the definition of $\xi$, Eq.~(\ref{paper1-n6}).

The solution of Eq.~(\ref{paper1-n5}) for the order parameter for $r \geq R$ is derived using the Taylor expansion of $\Delta \omega(m)$ about $m=m_b$, Eq.~(\ref{paper1-n1}).
In this expansion we have used up to the second-order term of
$(m-m_b)$ and have ignored the higher orders. We therefore need to
obtain the conditions under which we are allowed to use this
approximation. To do so we see that in Eq.~(\ref{paper1-n1}) by ignoring
terms of orders greater than $2$, we have assumed that these values
are less than the second-order term. The biggest of all ignored
terms is the third order one, we have therefore assumed that

\begin{equation} \label{paper1-n11}
\frac{\partial ^3\Delta\omega(m)}{\partial m^3} \bigg|_{m=m_b} \frac{(m-m_b)^3}{3!} \,\ll \,\frac{1}{2} \chi^{-1} (m-m_b)^2.
\end{equation} 
Using Eqs.~(\ref{paper1-4},\ref{paper1-n6}) in Eq.~(\ref{paper1-n11})
gives

\begin{equation}  \label{paper1-n12}
\frac{4b}{\kappa}\,m_b \,\xi^2(m(R)-m_b) \ll 1,
\end{equation}
as the requirement that must be satisfied. Here $m(R)$ is given by
Eq.~(\ref{paper1-n5}) at $r=R$.

\section{Results}     \label{results}

The free energy barriers to nucleation of the stable phase and of
the $+m$ phase are $\Delta F^*$ and $\Delta F^*_I$, respectively.
In each nucleation the formation of the new
phase occurs when the free energy barrier is of roughly $30kT$.
The figure $30kT$ is rather arbitrary \cite{Deb96} but the nucleation
rate is of order, number of molecules times $\exp(-\Delta F^*/kT)$
divided by the characteristic time scale of the solution. The
characteristic time scale of protein solutions is of order $1 \mu s$
\cite{MR97}. Thus a volume of the solution containing $10^9$ protein
molecules has a nucleation rate of order $10^9 \times e^{-30} \times
10^6 \approx 10s^{-1}$. Thus in a volume of the solution containing a
billion molecules, nuclei will appear in a fraction of a second. When
the barrier reaches $30kT$ the $-m$ phase is unstable and the
stable $+m$ phase nucleates immediately.
A barrier of about twice this, $60kT$, is required for the $-m$ phase
to be stable over long times, i.e., for it to be metastable.
Homogeneous nucleation of a new stable
phase near a second phase transition has
been studied before
\cite{TO98,tWF97,HD00,DZ00,PJPR00,Sear01a,Sear01b,Sear01c}. Here we
carried out numerical calculations for $\Delta m^*$ and $\Delta F^*$.
Near the critical point the scaling of the size of the critical
nucleus is \cite{Sear01a,Sear01b,Sear01c},

\begin{equation}  \label{paper1-n13}
\Delta m^* \sim |t|^{-\gamma} g_{\pm}\left(h/|t|^{\beta
\delta}\right), \end{equation}
%%%%%%%%%%%%%%%%%%%%%%%
and that of the free energy is

\begin{equation}  \label{paper1-e1}
\Delta F^*_s \sim |t|^\beta f_{\pm}\left(h/|t|^{\beta \delta}\right),
\end{equation}
%%%%%%%%%%%%%%%%%%%%%%%
for both temperatures above and below the critical temperature. Here
$\pm$ are related to $t>0$ and $t<0$, $\gamma$, $\beta$
and $\delta$ are the usual critical exponents, and take values of
$1$, $1/2$ and $3$ within mean-field theory.
In Eq.~(\ref{paper1-e1}) subscript $s$ stands for singular part of
the fringe. These results were predicted in earlier work
\cite{Sear01a,Sear01b,Sear01c} and we confirmed them by numerical
calculations. As the critical point is approached, the
magnetisation of the critical nucleus $\Delta m^*$ diverges, and so
does the derivative of the free energy barrier to nucleation $\Delta
F^*$, with respect to the temperature and to $h$. We are also
interested in the behaviour of $\Delta m^*$ and $\Delta F^*$ far from
the critical point.

\subsection{Far from the critical point}

\begin{figure}[t]
\begin{center}
\epsfig{file=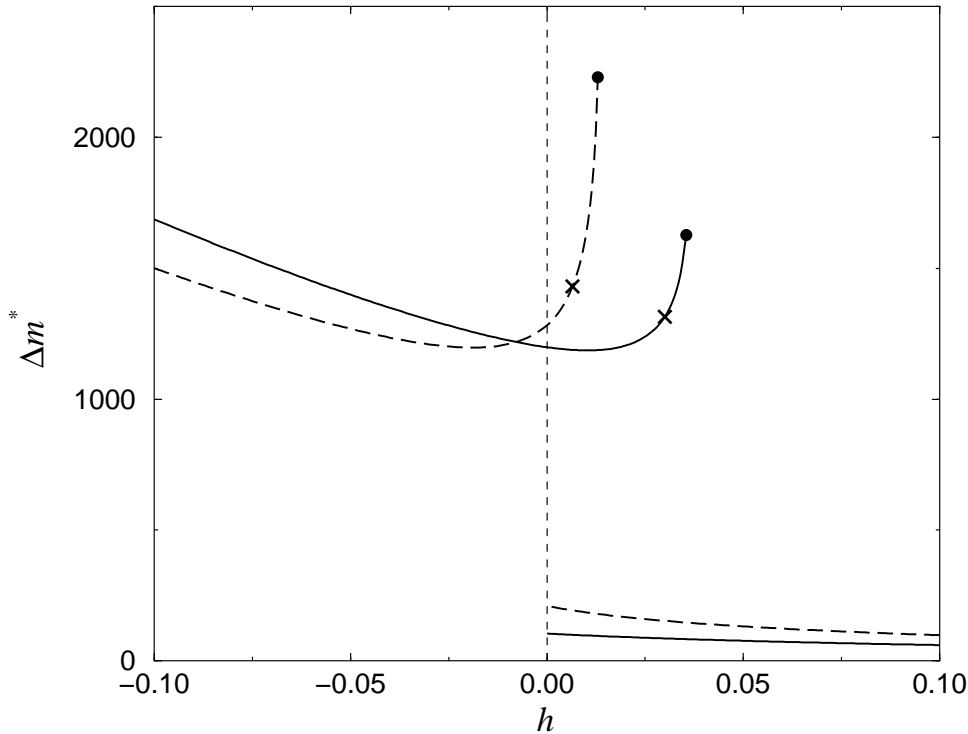,width=10cm,angle=-90}
\caption{$\Delta m^*$ versus $h$ at two temperatures below the
critical temperature, when $\kappa=1$ and $\Delta
\omega_{core}=-0.5$. Solid and long-dashed curves correspond to
$t=-0.5$ and $t=-0.25$, respectively.}
\label{paper1-figure1}
\end{center}
\end{figure}
%%%%%%%%%%%%%%%%%%
%%%%%%%%%%%%%%%%
\begin{figure}[t]
\begin{center}
\epsfig{file=
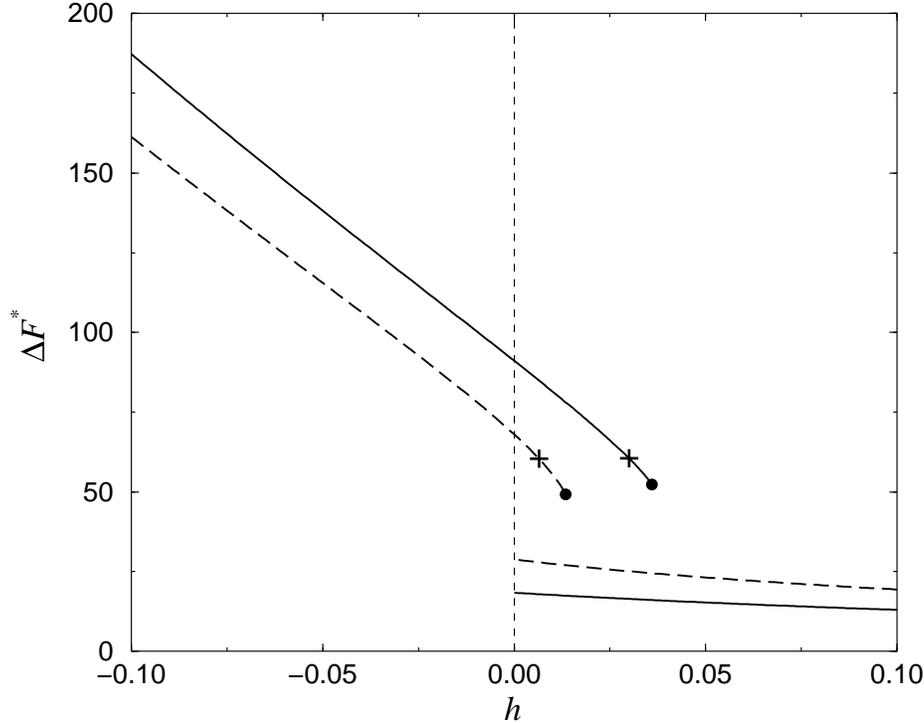,width=10cm,angle=-90}
\caption{$\Delta F^*$ versus $h$ for two different temperatures, when
$\kappa =1$ and $\Delta \omega_{core}=-0.5$. Solid and
long-dashed curves correspond to $t=-0.5$ and $t=-0.25$,
respectively.}
\label{paper1-figure2}
\end{center}
\end{figure}
%%%%%%%%%%%%%%%%%%
Here we study the total magnetisation and
the free energy of the critical nucleus,
$\Delta m^*$ and $\Delta F^*$, respectively, far from the critical
point --- $|t|$ not small.
In all the calculations we have fixed $a=b=m_c=1$.
In Figs.~\ref{paper1-figure1} and \ref{paper1-figure2} there are
graphs of $\Delta m^*$ and $\Delta F^*$, respectively, for two
different temperatures below and far from the critical temperature,
$t=-0.25$ and $t=-0.5$ with $\kappa=1$ and $\Delta
\omega_{core}=-0.5$. For each temperature, when $h<0$ the system is
in a state with a negative order parameter $m<0$ less than that at
the coexistence. When $h>0$ the system can be found in a state with
a positive order parameter $m>0$ greater than that at the
coexistence. For $h>0$ there are also states with $m<0$ between the
$-m$ branch of the coexistence curve and the spinodal. These states
are within two coexistence regions: the coexistence region between
$-m$ and $+m$ phases, and that between the equilibrium phases.
Therefore in both Figs.~\ref{paper1-figure1} and
\ref{paper1-figure2}, for each temperature there are values of $h>0$
for which $\Delta m^*$ and $\Delta F^*$ are double valued. These
values correspond to the $-m$ phase between the coexistence and
spinodal curves, and the $+m$ phase with a lower $\Delta m^*$ and
$\Delta F^*$.

In Fig.~\ref{paper1-figure1} for each temperature as $h$ increases
the curves on the right-hand side decrease. Those on the left-hand
side decrease first and then increase rapidly.
Decreasing $\Delta m^*$ is due to the fact that for a higher $h$, the
bulk order parameter $m_b$ is higher and thus it leads to a lower
excess order parameter $\Delta m^*$ in Eq.~(\ref{paper1-8.5}). However
the large and rapidly increasing values of $\Delta m^*$ are due to the
proximity to the spinodal.
As we approach the spinodal the response function of the order parameter,
$\chi$ defined as in Eq.~(\ref{paper1-n3}) diverges.
Therefore the total excess order parameter $\Delta m^*$ which scales as
$\chi$ \cite{Sear01a,Sear01b,Sear01c} will diverge as well.
As the temperature approaches $T_c$ ($|t|$ decreases),
$\Delta m^*$ diverges at a lower value of $h$.
This is because, as $T$ approaches $T_c$,
the gap between the coexistence and the spinodal curves decreases.
In Fig.~\ref{paper1-figure2} for each temperature as $h$ increases
$\Delta F^*$ decreases, due to the increase in the bulk order
parameter $m_b$. The phase with the higher order parameter
must overcome a lower free energy barrier $\Delta F^*$ to form a
nucleus of the equilibrium phase, whose order parameter we have set
equal to $1$.
Also in Fig.~\ref{paper1-figure2}, $\Delta F^*$ varies rapidly as
it approaches the spinodal. This is due to what is called the
nucleation theorem \cite{Kash82,VSR93,BMSSVR00} and states that the
derivative of the free energy barrier with respect to $h$ is equal to
minus the excess $m$, $\partial \Delta F^*/\partial h = -\Delta m^*$
\cite{Sear01a,Sear01b,Sear01c}. It shows that the larger $\Delta m^*$
is, the more rapidly the free energy barrier varies with $h$. As the
spinodal is approached $\Delta m^*$
diverges and thus $\Delta F^*$ drops rapidly with $h$.
When a transition from $-m$ to $+m$ phase occurs both $\Delta m^*$
and $\Delta F^*$ in Figs.~\ref{paper1-figure1} and
\ref{paper1-figure2} jump to lower values, because of the increase
in $m_b$. The height of this reduction is temperature-dependent
because the bulk order parameters $m_{co}=\pm \sqrt{\left(-t/4
\right)}$ are functions of $t=T-T_c$. As the temperature decreases,
the difference between these two order parameters increases, and
therefore the height of the drop in both $\Delta m^*$ and $\Delta
F^*$ increases.

For each temperature as $h$ increases $\Delta F^*_I$ decreases,
Eq.~(\ref{paper1-n14}), and for energy barriers of about $30kT$
a transition of $-m$ phase to $+m$ phase with lower values of
$\Delta m^*$ and $\Delta F^*$ occurs.
The cross marks on the graphs in Figs.~\ref{paper1-figure1} and
\ref{paper1-figure2} are the points where $\Delta F^*_I = 30kT$.
Also in the region with $h>0$ and $m<0$ the
solutions to $\Delta m^*$ and $\Delta F^*$ disappear at a value of
$h$ before the spinodal.
The circle marks on the graphs are the points for which the solutions
disappear.
For this value of $h$ in our numerical calculations the
solution to the order parameter profile of the critical nucleus,
$m^*(r)$, disappears. The solution to $m^*(r)$ is obtained by solving
Eq.~(\ref{paper1-8}) subject to the boundary condition
$m^*(r=r^*_c)=m_c$. In the region between the $-m$ branch of the
coexistence and the spinodal with $h>0$, Eq.~(\ref{paper1-8}) gives two
solutions for $m^*(r)$, out of which we choose the one with lower
free energy barrier, $\Delta F^*$. As $h$ increases, these two
solutions approach each other until they become equal at a specific
value of $h>0$. For any value of $h$ greater than this, then there is
no solution. This means that the critical nucleus
does not exist anymore and there is no well defined free
energy barrier to nucleation. As we see in Figs.~\ref{paper1-figure1}
and \ref{paper1-figure2} for each temperature the points with
$\Delta F^*_I=30kT$ (cross marks) are at the left side of those for
which the solutions disappear (circle marks). As for energy barriers
of $\Delta F^*_I=30kT$ a transition of $-m$ to $+m$ phase occurs,
therefore for these parameter values as the system approaches the
cross marks, the $-m$ phase will transform to $+m$ phase. Thus in the
experiments the states after the cross marks are not accessible.
As we see in Figs.~\ref{paper1-figure1} and \ref{paper1-figure2}
as the temperature approaches $T_c$ ($|t|$ decreases), the solution
disappears at a lower value of $h>0$. This is because as
$T$ approaches $T_c$, the coexistence and the spinodal curves
approach each other.
%%%%%%%%%%%%%%%%
\begin{figure}[t]
\begin{center}
\epsfig{file=
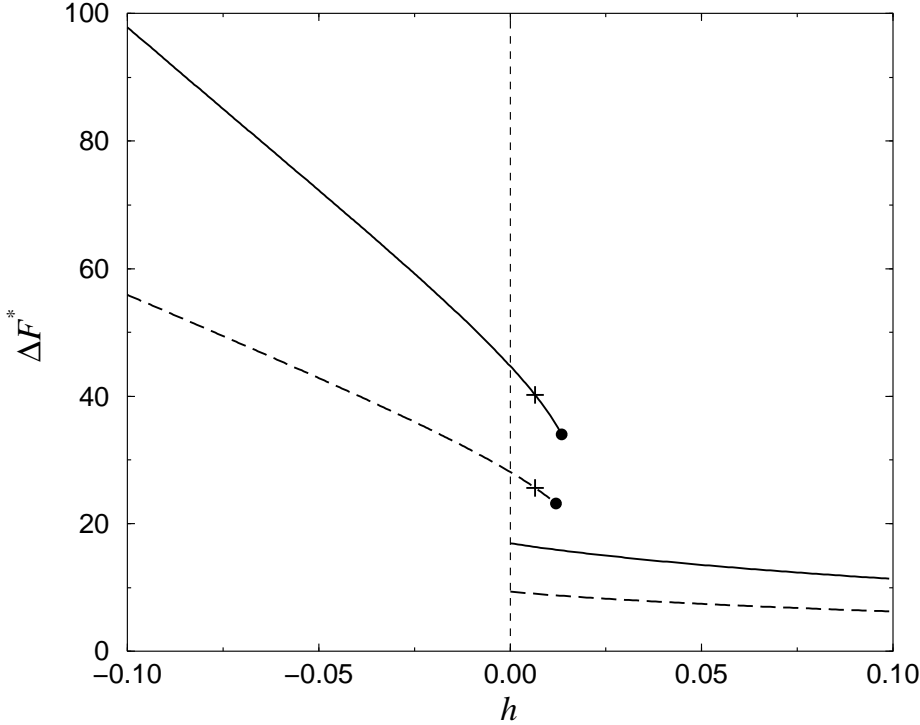,width=10cm,angle=-90}
\caption{$\Delta F^*$ versus $h$ for two $\Delta \omega_{core}$, when
$\kappa =1$ and $t=-0.25$. Solid and
long-dashed curves correspond to $\Delta \omega_{core}=-0.75$ and
$\Delta \omega_{core}=-1.25$, respectively.}
\label{paper1-figure3}
\end{center}
\end{figure}
%%%%%%%%%%%%%%%%%%

The stable phase can be attained via two processes \cite{Poon99}.
In one process $\Delta F^*$ drops below $30kT$ when $\Delta F^*_I$
is still large, then the stable phase is reached in one step through
nucleating in $-m$ phase.
In the second process $\Delta F^*_I$ drops below $30kT$ when $\Delta
F^*$ is still large, then the stable phase will be reached in
two steps: first the $+m$ phase nucleates in the $-m$ phase and grows
and then second, the stable phase nucleates in this $+m$ phase.
The fact that the nucleated phase is the one which has lower free
energy barrier and not necessarily the stable phase, is in accordance
with Ostwald's step rule.
In Fig.~\ref{paper1-figure2} with $\Delta \omega_{core}=-0.5$ the
barrier $\Delta F^*$ is always greater than $30kT$ when
$\Delta F^*_I$ drops below $30kT$. Thus on increasing $h$ the $+m$
phase nucleates before the equilibrium phase. The equilibrium phase
then nucleates in the $+m$ phase. But as $\Delta \omega_{core}<0$,
when $\Delta \omega_{core}$ becomes more negative then $\Delta F^*_c$
becomes more negative and therefore $\Delta F^*$ in
Eq.~(\ref{paper1-1}) drops. So for sufficiently negative
values of $\Delta \omega_{core}$, $\Delta F^*$ drops below $30kT$
before $\Delta F^*_I$ does and the stable phase nucleates before the
$+m$ phase.
To show this, in Fig.~\ref{paper1-figure3} graphs of $\Delta F^*$
are plotted for two values of $\Delta \omega_{core}=-0.75$
and $\Delta \omega_{core}=-1.25$, when $t=-0.25$.
Just as in the previous figures, the cross marks on each graph
correspond to where $\Delta F^*_I=30kT$ and the circle marks are
where the solutions disappear.
For $\Delta \omega_{core}=-0.75$, $\Delta F^*$ is greater than $30kT$
when $\Delta F^*_I$ drops below $30kT$. Thus the $+m$ phase nucleates
first and then the equilibrium phase nucleates in the $+m$ phase. For
$\Delta \omega_{core}=-1.25$ however $\Delta F^*$ drops below $30kT$
before $\Delta F^*_I$ does. Thus the equilibrium phase nucleates
before the $+m$ phase. The values of $h$ for which $\Delta
F^*_I=30kT$ (the cross marks) are independent of $\Delta
\omega_{core}$ and are the same on both graphs. This is because
$\Delta F^*_I$ in Eqs.~(\ref{paper1-n14},\ref{paper1-n15})
depends only on the temperature and on $h$, and is independent of
$\Delta \omega_{core}$.

%%%%%%%%%%%%%%%%
\begin{figure}[t]
\begin{center}
\epsfig{file=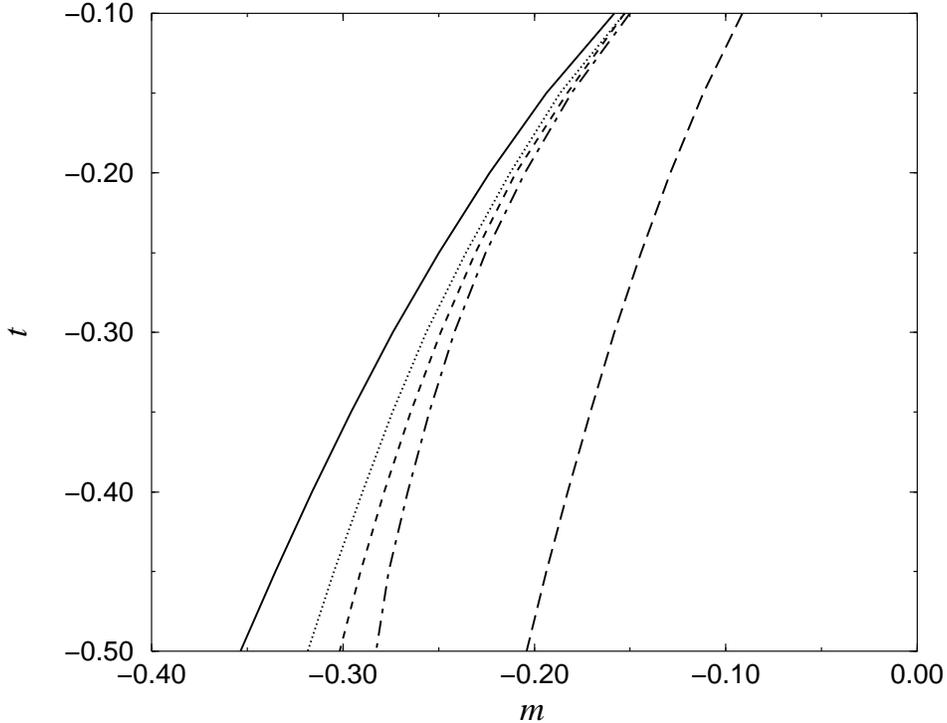,width=10cm,angle=-90}
\caption{Phase diagram for $t$ versus $m$. Solid and long-dashed
lines are coexistence and spinodal, respectively. The others
correspond to $\Delta F^*_I = 30kT$, dotted, dashed and dot-dashed
lines are for $\kappa=1,1.5$ and $2$, respectively.}
\label{paper1-figure4}
\end{center}
\end{figure}
%%%%%%%%%%%%%%%%%%
%%%%%%%%%%%%%%%%
\begin{figure}[t]
\begin{center}
\epsfig{file=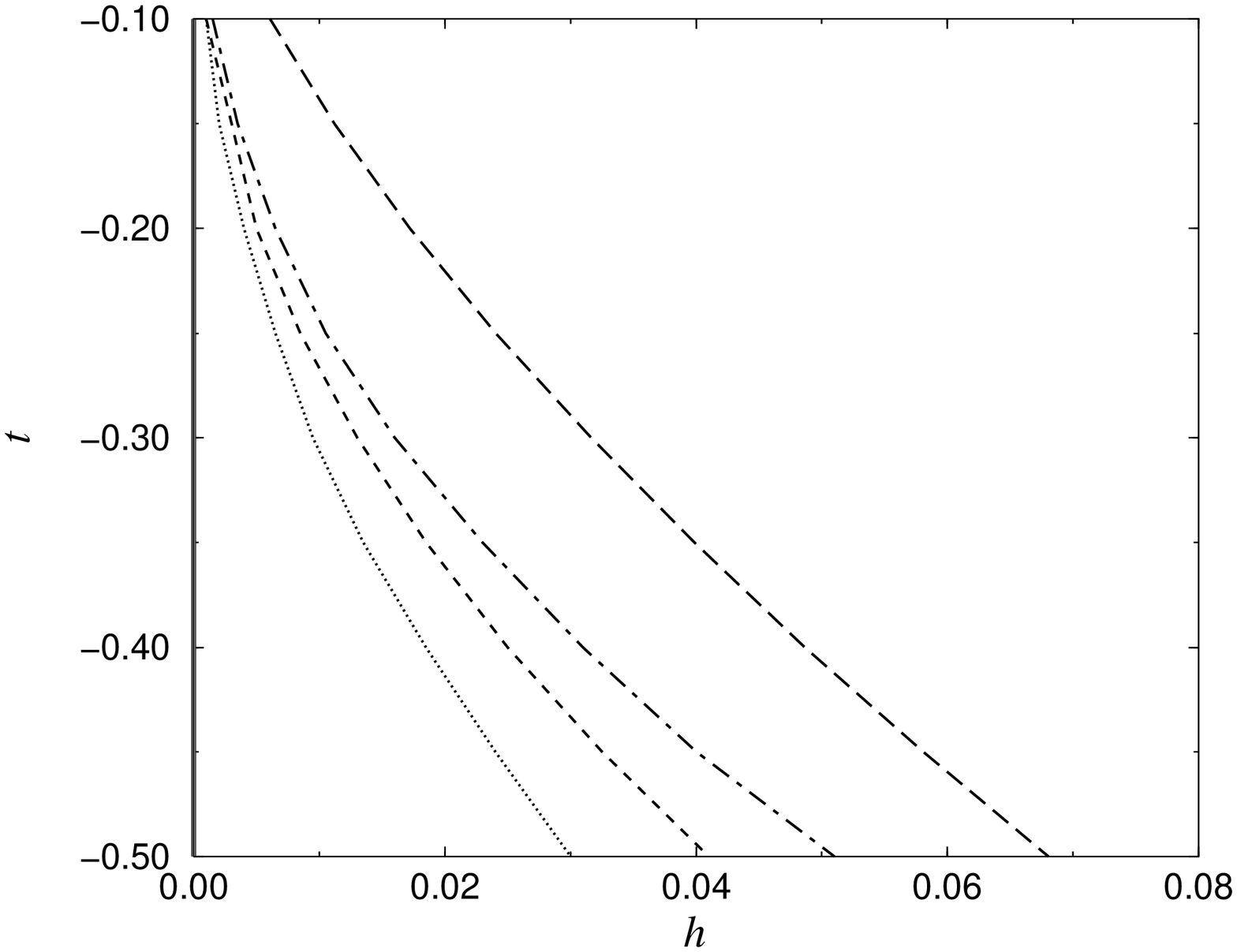,width=10cm,angle=-90}
\caption{Phase diagram for $t$ versus $h$. Coexistence lies along the
$t$ axis. Long-dashed curve is the spinodal. The others correspond to
$\Delta F^*_I = 30kT$, dotted, dashed and dot-dashed lines are for
$\kappa=1,1.5$ and $2$, respectively.}
\label{paper1-figure5}
\end{center}
\end{figure}
%%%%%%%%%%%%%%%%%%

The two step route to the equilibrium phase that we find is not
dissimilar to that proposed for the crystallisation of some polymer
melts by Olmsted et.~al.~\cite{OPMTR98}. They study a system with the
same phase diagram as considered here but their equivalent of the
$-m$ to $+m$ phase transition proceeds via spinodal decomposition
and not nucleation as here.
Thus the results here apply to liquids and colloidal suspensions,
where the new phase nucleates, whereas Olmsted et.~al.'s approach is
more applicable to polymers.

We are also interested to study where the free energy barrier
$\Delta F^*_I = 30kT$,
and a nucleation from the $-m$ phase to the $+m$ phase occurs.
Up to now we considered systems with $\kappa=1$.
We also like to know the effect of increasing $\kappa$
on the calculations.
In Figs.~\ref{paper1-figure4} and \ref{paper1-figure5}
there are diagrams of $t$ versus $m$, and $t$
versus $h$, respectively for three different values of $\kappa$ when
$\Delta \omega_{core}=-0.5$. In Fig.~\ref{paper1-figure4} the
solid (outermost) and long-dashed (innermost) curves represent the
coexistence and the spinodal, respectively.
The other curves in Fig.~\ref{paper1-figure4} between
the coexistence and spinodal curves correspond to the points
with energy barriers of $\Delta F^*_I = 30kT$ for different
$\kappa$. From left to right, dotted, dashed and dot-dashed
curves are for $\kappa=1, 1.5$ and $2$, respectively.
In Fig.~\ref{paper1-figure5} the solid line $h=0$ along
axis $t$ corresponds to coexistence, and the long-dashed curve
(the outermost) represents spinodal.
The other curves in Fig.~\ref{paper1-figure5} between
the coexistence and the spinodal correspond to the points
with $\Delta F^*_I = 30kT$ for different $\kappa$.
From left to right, dotted, dashed and dot-dashed curves
represents $\kappa=1, 1.5$ and $2$, respectively.
As $\kappa$ increases there is a shift towards larger
values of both $m$ and $h$ for $\Delta F^*_I = 30kT$.
This shift can be explained as follows.
For a fixed temperature and external field $h$,
a higher $\kappa$ yields a higher free energy barrier to nucleation
of the $+m$ phase $\Delta F^*_I$
(\ref{paper1-n14},\ref{paper1-n15}). Now for each fixed temperature
and $\kappa$, as $h$ increases the energy barrier $\Delta F^*_I$,
Eq.~(\ref{paper1-n14}), decreases until it reaches energies of about
$30kT$. Therefore at each fixed temperature, for a higher $\kappa$
the nucleation occurs at a higher external field $h$
and consequently at a higher order parameter $m$.
Increasing $\kappa$ also increases $\Delta F^*$ Eq.~(\ref{paper1-1}),
thus increasing $\kappa$ pushes nucleation to both
the $+m$ and the stable phase to larger values of $h$.

\section{Conclusions}

Homogeneous nucleation of a new phase at a first-order transition
near a second transition is considered. The second transition is an
Ising-like or a vapour-liquid-like transition, and lies
within the coexistence region of the equilibrium transition.
Numerical calculations were performed to
calculate the excess order parameter $\Delta m^*$ and the free energy
barrier to nucleation $\Delta F^*$, of the critical nucleus.
For the states in the negative magnetisation phase with $h>0$
the system is within the coexistence regions of both transitions.
Then in these states either the positive magnetisation phase
or the equilibrium phase can nucleate.
The free energy barriers to nucleation are equal to $\Delta F^*_I$
and $\Delta F^*$, for the former and the latter, respectively.
The phase which nucleates is the one with the lower free energy
barrier. This is in accordance with Ostwald's step rule, which
states that the nucleated phase is not necessarily the stable phase,
but is the one with the lowest free energy barrier to nucleation.
The stable phase can be attained via two processes.
In the first process  $\Delta F^*$ becomes small enough to allow
nucleation when $\Delta F^*_I$ is still large, then the stable phase
nucleates.
In the second process $\Delta F^*_I$ becomes small enough to allow
nucleation when $\Delta
F^*$ is still large, then the stable phase will be reached in
two steps: first the positive magnetisation phase nucleates and
grows and then second, the stable phase nucleates
in this positive magnetisation phase.

When the Ising-like phase transition occurs, both $\Delta m^*$
and $\Delta F^*$ jump to lower values, due to the increase in the
bulk order parameter.
As the bulk order parameter at coexistence is a function of the
temperature, then the height of the jump is temperature-dependent.
For a lower temperature, the jump in the order parameter increases
and therefore the jumps in $\Delta m^*$ and the barrier to
nucleation $\Delta F^*$ increase.
This may mean a jump from a nucleation barrier too high to permit
significant nucleation, to a very small barrier that results in rapid
nucleation with many nuclei forming.
This effect may cause problems for crystallographers wanting to
achieve a small but non-zero rate of nucleation in order to obtain a
few large crystals.
We also observed that as the spinodal is
approached, $\Delta m^*$ diverges and the free energy barrier $\Delta
F^*$ varies rapidly with $h$.

\vskip 0.35cm
\noindent
{\large \bf Acknowledgements}

\vskip 0.15cm
\noindent
This work is supported by EPSRC (GR/N36981).

%\newpage

\bibliographystyle{statis}
\bibliography{phatra}

\begin{thebibliography}{10}

\bibitem{Deb96}
P.~G. Debenedetti, {\it Metastable Liquids\/} (Princeton University Press,
  Princeton, 1996).

\bibitem{BBPOB91}
M.~L. Broide, C.~R. Berland, J.~Pande, O.~O. Ogun and G.~B. Benedek, Proc. Nat.
  Acad. Sci. {\bf 88}, 5660 (1991).

\bibitem{MR97}
M.~Muschol and F.~Rosenberger, J. Chem. Phys. {\bf 107}, 1953 (1997).

\bibitem{RVMT96}
F.~Rosenberger, P.~G. Vekilov, M.~Muschol and B.~R. Thomas, J. Crys. Growth
  {\bf 168}, 01 (1996).

\bibitem{CH98}
N.~Chayen and J.~Helliwell, Physics World, May {\bf 43} (1998).

\bibitem{Piaz00}
R.~Piazza, Curr. Opin. Coll. Int. Sci. {\bf 5}, 38 (2000).

\bibitem{GW94}
A.~George and W.~Wilson, Acta Crystallogr. D {\bf 50}, 361 (1994).

\bibitem{RZZ96}
D.~Rosenbaum, P.~C. Zamora and C.~F. Zukoski, Phys. Rev. Lett. {\bf 76}, 150
  (1996).

\bibitem{TO98}
V.~Talanquer and D.~W. Oxtoby, J. Chem. Phys. {\bf 109}, 223 (1998).

\bibitem{tWF97}
P.~R. ten Wolde and D.~Frenkel, Science {\bf 277}, 1975 (1997).

\bibitem{HD00}
C.~Haas and J.~Drenth, J. Phys. Chem. B {\bf 104}, 368 (2000).

\bibitem{DZ00}
N.~M. Dixit and C.~F. Zukoski, J. Coll. Int. Sci. {\bf 228}, 359 (2000).

\bibitem{PJPR00}
D.~Pini, G.~Jialin, A.~Parola and L.~Reatto, Chem. Phys. Lett. {\bf 327}, 209
  (2000).

\bibitem{Sear01a}
R.~P. Sear, Phys. Rev. E {\bf 63}, 066105 (2001).

\bibitem{Sear01b}
R.~P. Sear, J. Chem. Phys. {\bf 114}, 3170 (2001).

\bibitem{Sear01c}
R.~P. Sear, submitted to J. Chem. Phys .

\bibitem{Ost97}
W.~Ostwald, Z. Phys. Chem. {\bf 22}, 289 (1897).

\bibitem{tWF99}
P.~R. ten Wolde and D.~Frenkel, Phys. Chem. Chem. Phys. {\bf 1}, 2191 (1999).

\bibitem{ST33}
I.~N. Stranski and D.~Totomanow, Z. Phys. Chem. {\bf 163}, 399 (1933).

\bibitem{Abr74}
F.~F. Abraham, {\it Homogeneous Nucleation Theory. The Pretransition Theory of
  Vapor Condensation\/} (Academic Press, New York, 1974).

\bibitem{VW26}
M.~Volmer and A.~Weber, Z. Phys. Chem. {\bf 119}, 277 (1926).

\bibitem{CH59}
J.~W. Cahn and J.~E. Hilliard, J. Chem. Phys. {\bf 31}, 688 (1959).

\bibitem{CH58}
J.~W. Cahn and J.~E. Hilliard, J. Chem. Phys. {\bf 28}, 258 (1958).

\bibitem{Wid85}
B.~Widom, Chem. Soc. Rev. {\bf 14}, 121 (1985).

\bibitem{Kad00}
L.~P. Kadanoff, {\it Statistical Physics\/} (World Scientific, Singapore,
  2000).

\bibitem{CL95}
P.~M. Chaikin and T.~C. Lubensky, {\it Principles of Condensed Matter
  Physics\/} (Cambridge University Press, Cambridge, 1995).

\bibitem{LL80}
L.~D. Landau and E.~M. Lifshitz, {\it Statistical Physics, Part 1\/}, vol.~5 of
  {\it Course of Theoretical Physics\/}, 3rd edn. (Pergamon Press Ltd., Oxford,
  1980).

\bibitem{Bin84}
K.~Binder, Phys. Rev. A {\bf 29}, 341 (1984).

\bibitem{PTVF92}
W.~H. Press, S.~A. Teukolsky, W.~T. Vetterling and B.~P. Flannery, {\it
  Numerical Recipes in {C}\/} (Cambridge University Press, Cambridge, 1992).

\bibitem{Kash82}
D.~Kashchiev, J. Chem. Phys. {\bf 76}, 5098 (1982).

\bibitem{VSR93}
Y.~Viisanen, R.~Strey and H.~Reiss, J. Chem. Phys. {\bf 99}, 4680 (1993).

\bibitem{BMSSVR00}
R.~K. Bowles, R.~McGraw, P.~Schaaf, B.~Senger, J.~C. Voegel and H.~Reiss, J.
  Chem. Phys. {\bf 113}, 4524 (2000).

\bibitem{Poon99}
W.~C.~K. Poon, F.~Renth, R.~M.~L. Evans, D.~J. Fairhurst, M.~E. Cates and P.~N.
  Pusey, Phys. Rev. Lett. {\bf 83}, 1239 (1999).

\bibitem{OPMTR98}
P.~D. Olmsted, W.~C.~K. Poon, T.~C.~B. Mcleish, N.~J. Terrill and A.~J. Ryan,
  Phys. Rev. Lett. {\bf 81}, 373 (1998).

\end{thebibliography}

\end{document}